\documentclass[a4paper]{jpconf}
\usepackage{graphics}
\usepackage{bbm}
\usepackage{amsmath}

\def\colora{}
\def\colorb{}

\def\colord{}

\def\empile#1\over#2{\mathrel{\mathop{\kern 0pt#1}\limits_{#2}}}
\newcommand{\slvarepsilon}{\raise.15ex\hbox{$/$}\kern-.53em\hbox{$\varepsilon$}}
\newcommand{\slL}{\raise.15ex\hbox{$/$}\kern-.53em\hbox{$L$}}
\newcommand{\slP}{\raise.15ex\hbox{$/$}\kern-.53em\hbox{$P$}}
\newcommand{\slD}{\raise.15ex\hbox{$/$}\kern-.53em\hbox{$D$}}
\newcommand{\slp}{\raise.1ex\hbox{$/$}\kern-.63em\hbox{$p$}}
\newcommand{\slq}{\raise.1ex\hbox{$/$}\kern-.53em\hbox{$q$}}
\newcommand{\slv}{\raise.1ex\hbox{$/$}\kern-.63em\hbox{$v$}}
\newcommand{\slR}{\raise.15ex\hbox{$/$}\kern-.53em\hbox{$R$}}
\newcommand{\slQ}{\raise.15ex\hbox{$/$}\kern-.53em\hbox{$Q$}}
\newcommand{\slK}{\raise.15ex\hbox{$/$}\kern-.53em\hbox{$K$}}
\newcommand{\slk}{\raise.15ex\hbox{$/$}\kern-.53em\hbox{$k$}}
\newcommand{\slSigma}{\raise.15ex\hbox{$/$}\kern-.53em\hbox{$\Sigma$}}
\newcommand{\slcalP}{\raise.15ex\hbox{$/$}\kern-.63em\hbox{$\cal P$}}
\newcommand{\slcalA}{\raise.15ex\hbox{$/$}\kern-.63em\hbox{$\cal A$}}
\newcommand{\slA}{\raise.15ex\hbox{$/$}\kern-.73em\hbox{$A$}}
\newcommand{\slbfA}{\raise.15ex\hbox{$/$}\kern-.73em\hbox{${\imb A}$}}
\newcommand{\slpartial}{\raise.15ex\hbox{$/$}\kern-.53em\hbox{$\partial$}}
\newcommand{\sla}{\raise.15ex\hbox{$/$}\kern-.53em\hbox{$a$}}
\newcommand{\slb}{\raise.15ex\hbox{$/$}\kern-.53em\hbox{$b$}}
\newcommand{\slc}{\raise.15ex\hbox{$/$}\kern-.53em\hbox{$c$}}
\newcommand{\slC}{\raise.15ex\hbox{$/$}\kern-.63em\hbox{$C$}}

\def\p{{\boldsymbol p}}

\def\x{{\boldsymbol x}}

\def\u{{\boldsymbol u}}
\def\v{{\boldsymbol v}}

\def\bs{\boldsymbol}

\begin{document}
\title{The early stages of a high energy heavy ion collision}

\author{Fran\c{c}ois Gelis}

\address{Institut de Physique Th\'eorique, CEA/DSM/Saclay, 91191 Gif sur Yvette cedex, France}

\ead{francois.gelis@cea.fr}

\begin{abstract}
  At high energy, the gluon distribution in nuclei reaches large
  densities and eventually saturates due to recombinations, which
  plays an important role in heavy ion collisions at RHIC and the
  LHC. The Color Glass Condensate provides a framework for resumming
  these effects in the calculation of observables. In this talk,
  I present its application to the description of the early stages of
  heavy ion collisions.
\end{abstract}

\section{Heavy ion collisions and color glass condensate}
\begin{figure}[h]
\begin{center}
\resizebox*{!}{4.5cm}{\includegraphics{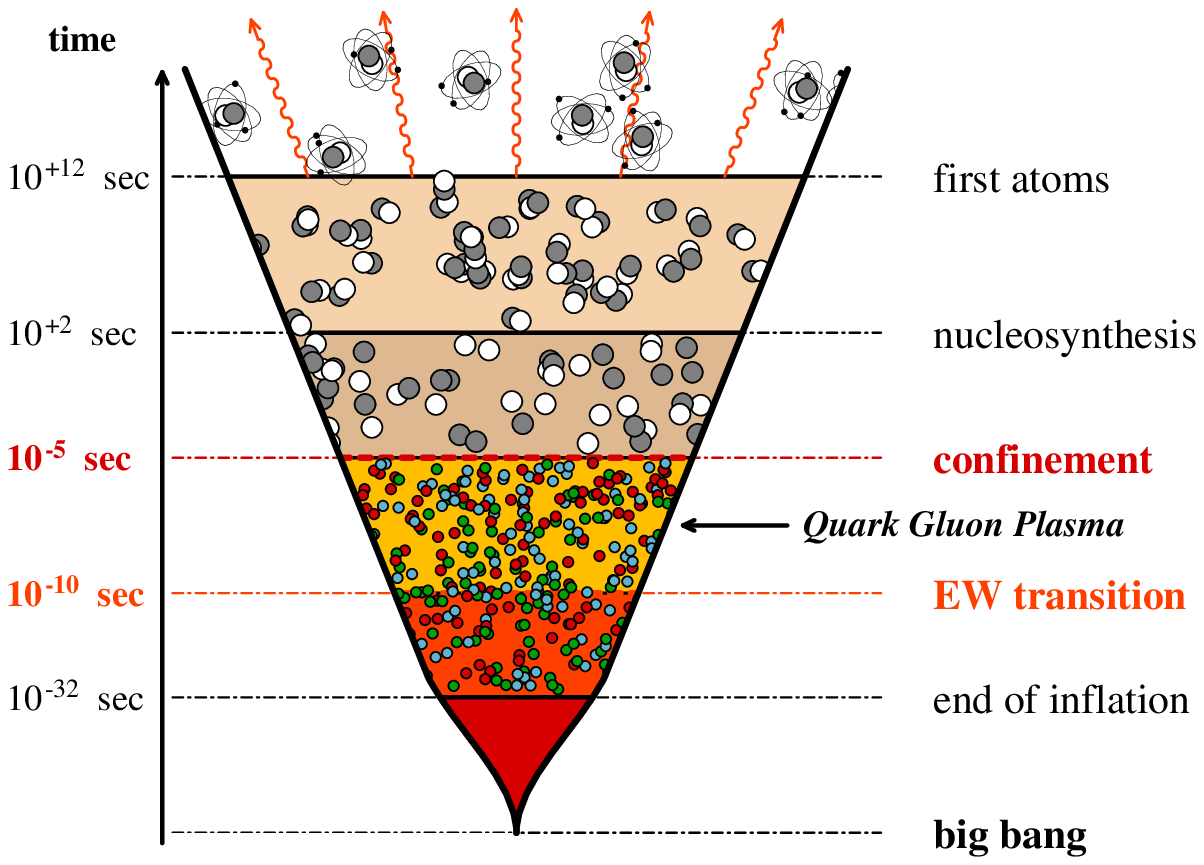}}
\hfil
\resizebox*{!}{4.5cm}{\includegraphics{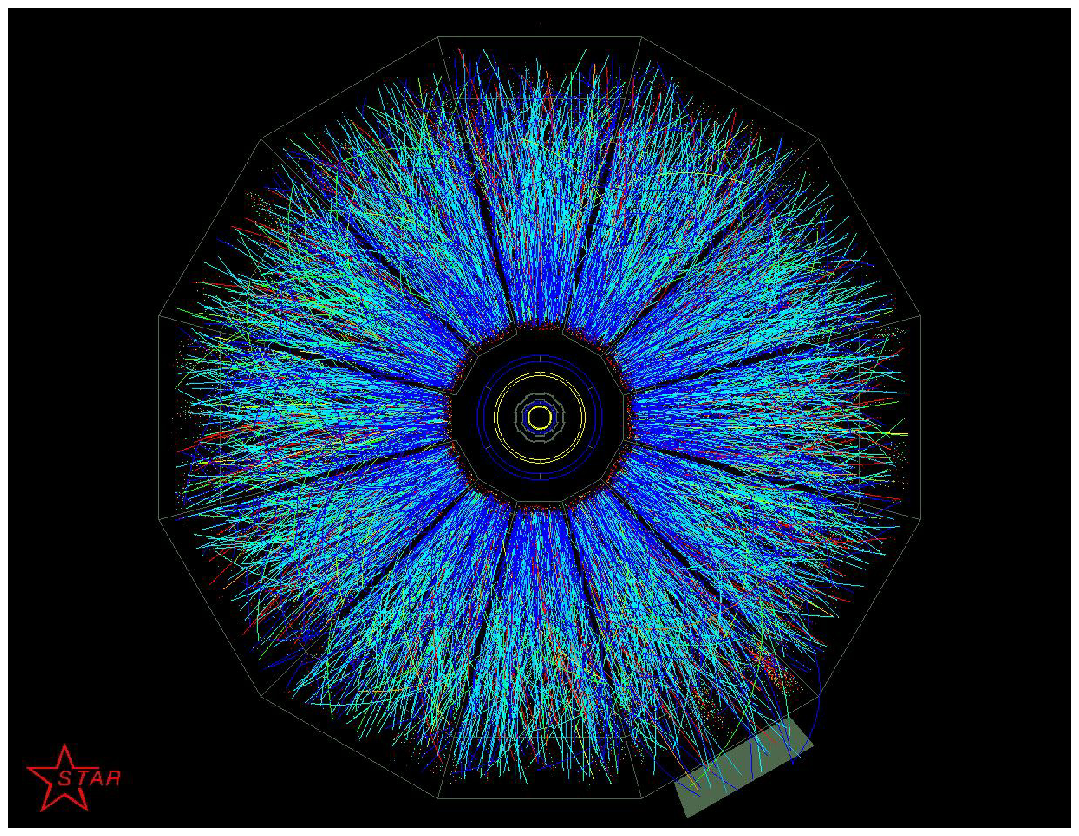}}
\end{center}
\caption{\label{fig:bigbang}Left: the quark gluon plasma in the early
  universe. Right: event display of a heavy ion collision at RHIC
  (STAR collaboration).}
\end{figure}

Because of the asymptotic freedom of strong interactions, the quark
gluon plasma is the expected state of nuclear matter at sufficiently
high temperature (above $T\sim 170$~MeV). Therefore, the early
universe was presumably made of free quarks and gluons at times
shorter than $\tau\sim 10^{-5}$~seconds (figure \ref{fig:bigbang},
left). Unfortunately, it seems that the confinement transition is too
weak to have left any imprint that could be observable nowadays via
astronomical observations, and we must therefore seek ways to
study this state of nuclear matter in the laboratory.
It is expected that the conditions of this transition can be reached
in collisions of large nuclei at ultra-relativistic energies. Such
experiments are ongoing at RHIC (Brookhaven, USA) and at the the LHC
(CERN), respectively at a maximum center of mass energy of $200$~GeV
and $2.76~$TeV per nucleon pair (see the figure \ref{fig:bigbang},
right).  There is ample evidence that the energy density reached in
these collisions is (for a brief period of time) far above the
critical energy density, and that deconfined quark-gluon matter is
produced.

\begin{figure}[h]
\begin{center}
\resizebox*{!}{4cm}{\includegraphics{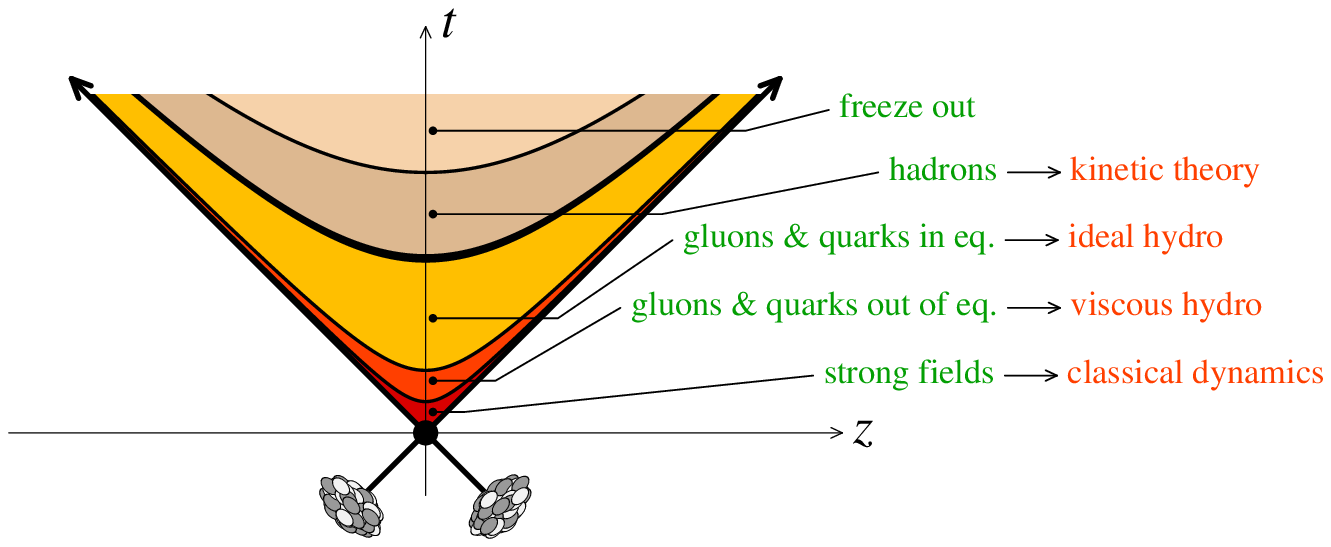}}
\hfil
\resizebox*{!}{4.5cm}{\includegraphics{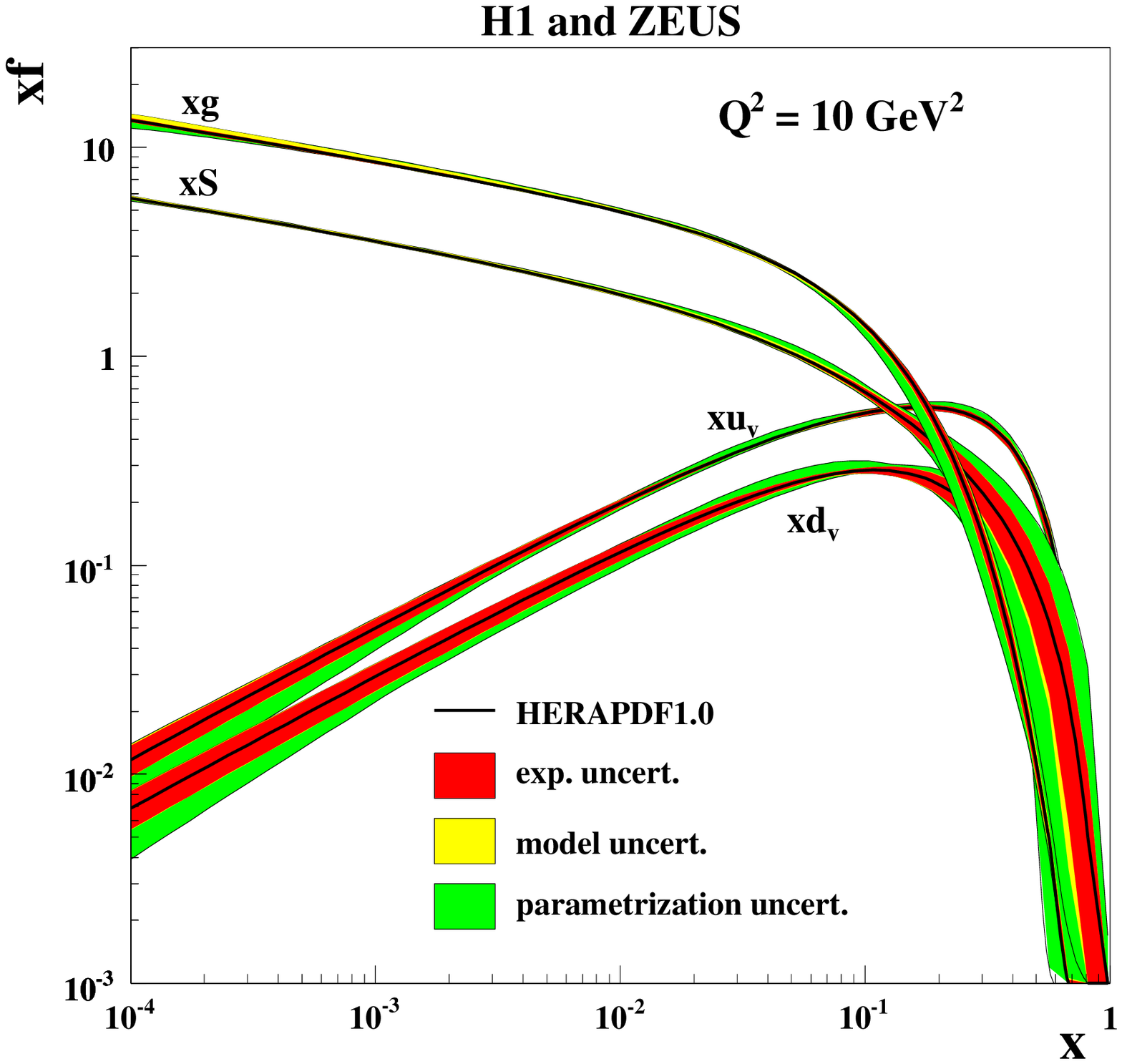}}
\end{center}
\caption{\label{fig:stages}Left: stages of a heavy ion
  collision. Right: proton parton distributions (from \cite{Aarona2}).}
\end{figure}

\begin{figure}[h]
\begin{center}
\resizebox*{!}{4.5cm}{\includegraphics{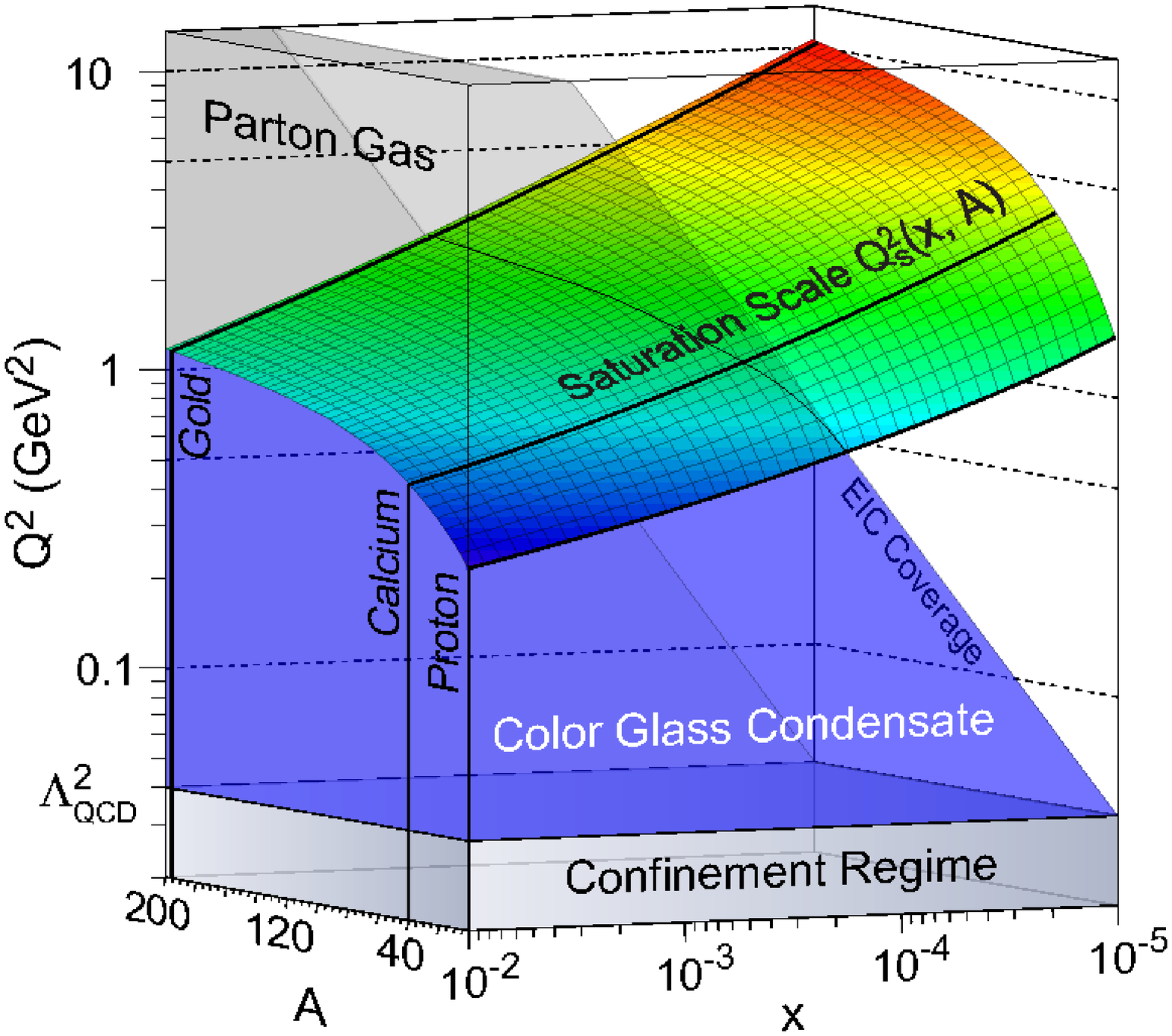}}
\hfil
\resizebox*{!}{3.7cm}{\includegraphics{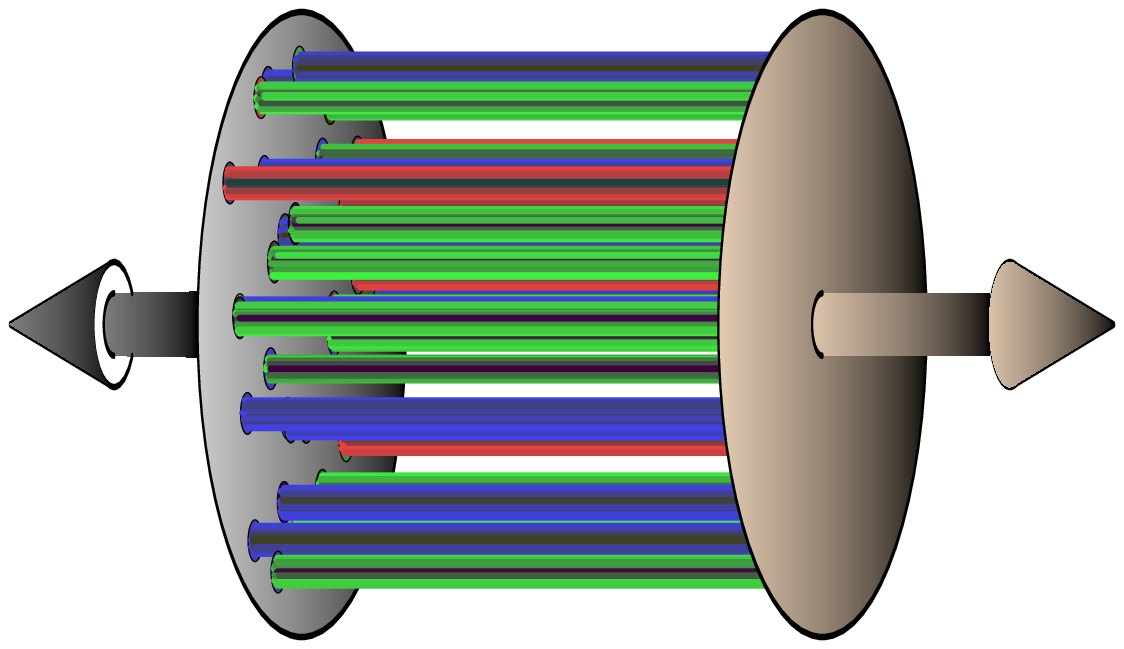}}
\end{center}
\caption{\label{fig:qsat}Left: saturation momentum as a function of
  $x$ and mass number $A$ (from \cite{DeshpEM1}). Right: color flux
  tubes.}
\end{figure}

From a theoretical perspective, the evolution of the matter produced
in the collision can be divided in several stages, that are handled by
different tools (figure \ref{fig:stages}, left). 
Broadly speaking, there are three main stages: (i) the collision
itself, and a brief period of time afterwards, (ii) a longer stage
where the matter is close to local thermal equilibrium, and (iii) the
final stage during which the matter has become too dilute to remain in
equilibrium. The focus of this talk is on the first of these three
stages. During the time evolution, the matter produced in the
collision expands, and therefore cools down. Therefore, this first
stage is also characterized by the highest momentum scales which, by
virtue of asymptotic freedom, suggests that one could describe it by a
weak coupling expansion in Quantum Chromo-Dynamics
(QCD). Unfortunately, the simplicity of the QCD Lagrangian,
\begin{equation}
{\cal L}=-\frac{1}{4}{\colorb F^2} + \sum_{\rm flavors} {\colora\overline\psi_f}(i{\colorb\slD}-{\colora m_f}){\colora\psi_f}\; ,
\end{equation}
is deceptive and the situation is complicated by the very large gluon
density in a nucleus at the relevant scales. In heavy ion collisions,
the partons that play a role in the production of the bulk of the
final state carry a very small fraction $x$ of the momentum of their
parent nucleon, $x\sim P_\perp/\sqrt{s}$ (where $P_\perp$ is the
transverse momentum of a typical final state particle and $\sqrt{s}$
the center of mass energy per nucleon pair.)  This value ranges from
$x\sim 10^{-2}$ (at RHIC) to $x\sim 10^{-4}$ (at the
LHC). Measurements of the parton distributions in a proton (at HERA
for instance, see the right panel of the figure \ref{fig:stages}) show
that at such low values of $x$, the nucleon content is predominantly
made of gluons, that have a very large density. This increase of the
gluon density at low $x$ is due to the large emission probability of
soft gluons (the differential probability behaves like $dP\sim
\alpha_s dx/x$, where $\alpha_s$ is the coupling constant of strong
interactions). However, this increase cannot continue forever: indeed,
when the gluon occupation number becomes of the order of the inverse
coupling $\alpha_s^{-1}$, the reverse process has a high probability
-- two gluons can then recombine, leading to a saturation of the gluon
density~\cite{GriboLR1}. This non-linear saturation mechanism
generates a dynamical momentum scale --the saturation momentum,
denoted $Q_s$--, below which saturation effects are important. This
scale increases at small momentum fraction $x$ and increases with the
mass number $A$ of a nucleus, as shown in the left panel of the figure
\ref{fig:qsat}. This is why saturation is expected to play a more
important role in heavy ion collisions than in proton-proton
collisions, even more so at the energy of the LHC.

\begin{figure}[h]
\begin{center}
\resizebox*{!}{3.4cm}{\includegraphics{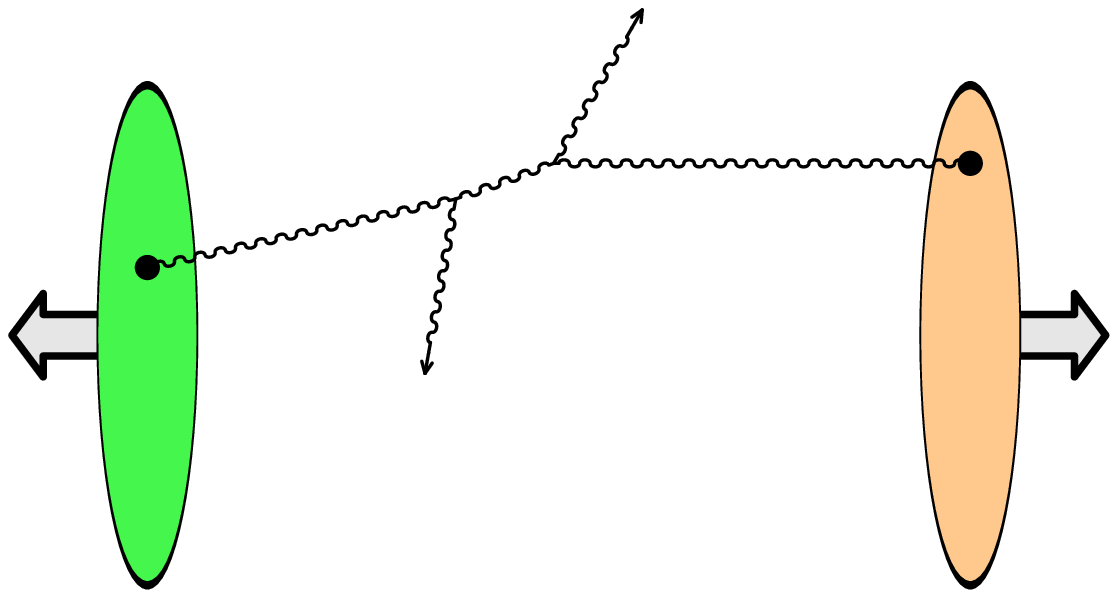}}
\hfil
\resizebox*{!}{3.4cm}{\includegraphics{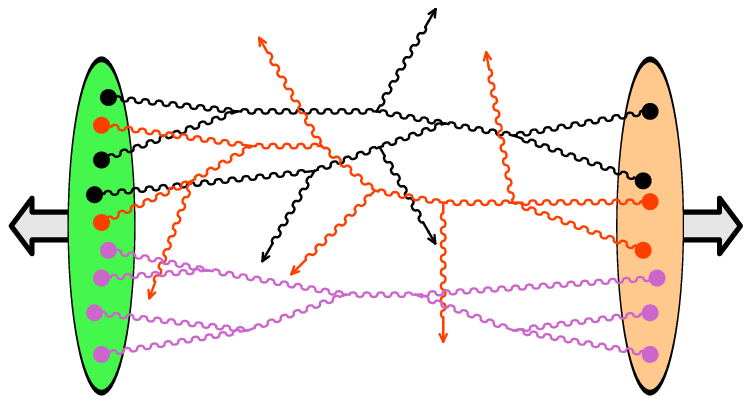}}
\end{center}
\caption{\label{fig:dilute}Left: collision of two dilute projectiles. Right: collision of two densely occupied projectiles.}
\end{figure}

Gluon densities of order $\alpha_s^{-1}$ lead to a breakdown
of the plain perturbative expansion, making the saturation regime
non-perturbative. For instance, Feynman graphs such as the one on the
right of the figure \ref{fig:dilute} are as large as the one on the
left.
The resummation of the infinite series of relevant graphs would be
extremely cumbersome if attacked via standard Feynman rules. The Color
Glass Condensate (CGC) \cite{IancuLM3} is an
effective theory based on QCD that reorganizes the perturbative
expansion in order to simplify this resummation.  It does so by
neglecting the quarks and by treating the fast partons (those at large
$x$) as static color sources $J$ moving along the trajectories of the
two projectiles, via the following effective Lagrangian
\begin{equation}
{\cal S}=\underbrace{-\frac{1}{4}\int F_{\mu\nu}F^{\mu\nu}}_{{\cal S}_{_{\rm YM}}}
+\int\underbrace{\vphantom{\int}({\colord J^\mu_1+J^\mu_2})}_{\mbox{fast partons}} {\colorb A_\mu}\; .
\end{equation}
In the saturated regime, these sources $J_{1,2}$ are large and must be
resummed to all orders. One can write an expansion in powers of the
coupling for observables, e.g.
\begin{equation}
\frac{dN_1}{dy d^2\vec\p_\perp}\sim {\colorb\frac{1}{\alpha_s}}\;
\Big[c_0+c_1\,{\colord \alpha_s}+c_2\,{\colord \alpha_s^2}+\cdots\Big]
\end{equation}
for the single inclusive gluon spectrum, but the coefficients $c_i$ in
this expansion are non-perturbative functions of $J$. However, the
advantage of this formulation is that quantum fields driven by a large
source behave classically in a first approximation. This implies that at
Leading Order in $\alpha_s$, observables can be obtained by solving
the classical Yang-Mills equations \cite{KrasnV2},
\begin{equation}
\big[{\colord{\cal D}_\mu},{\colord{\cal F}^{\mu\nu}}\big]={\colorb J_1^\nu}+{\colorb J_2^\nu}\quad,\quad
\lim_{t\to -\infty}{\colord{\cal A}^\mu(t,\vec\x)}=0\; ,
\end{equation}
instead of computing an infinite series of Feynman graphs.
It turns out that this classical solution has a very peculiar feature,
imposed by the geometry of the collision: at a proper time $\tau=0^+$,
the lines of the chromo-electric and chromo-magnetic fields are all
parallel to the collision axis \cite{LappiM1} (see the figure
\ref{fig:qsat}, right). These fields form color flux tubes in the $z$
direction, that have a typical transverse size of order $Q_s^{-1}$.  And
the energy-momentum tensor of such a configuration of color fields is
no less peculiar, since its longitudinal pressure is negative. Thus,
this matter is initially very far from local thermodynamical
equilibrium, and it is in fact a long standing puzzle to understand how
(and if) it evolves towards thermalization.

\section{Initial state factorization}
Before we turn to the question of thermalization, let us first discuss
another important issue: how do we know what sources $J_{1,2}$ we
should use in the CGC approach to heavy ion collisions? The short
answer is that for a given collision, we don't. $J$ reflects the
configuration (color and position in the transverse plane) of the fast
partons of a nucleus at the time it collides with the second nucleus,
and therefore the best we can hope for is to know the probability of
having a certain configuration. Thus, the original question translates
into: how do we know this distribution? It is possible to constrain
this distribution by measurements done in other experiments, such as
Deep Inelastic Scattering (DIS) on nuclei, provided one justifies its
universality (i.e. that the distribution is the same for all inclusive
observables in all experiments). In the case of DIS, it is know that
the relevant distribution obeys the JIMWLK equation, that controls its
evolution with energy.

In order to check whether the same is true in heavy ion collisions,
one needs to consider higher order corrections to inclusive
observables. These corrections contain logarithms of the collision
energy, and one should check that these logarithms are the same as in
DIS. At NLO, the single inclusive gluon spectrum can be written in the
following way \cite{GelisLV3},
\begin{equation}
\left.{\colorb\frac{dN_1}{dyd^2\vec\p_\perp}}\right|_{_{\rm NLO}}
=
\Bigg[
\frac{1}{2}\int\limits_{_{\vec\u,\vec\v\in\Sigma}}
{\colorb{\cal G}(\vec\u,\vec\v)}\,{\mathbbm T}_\u{\mathbbm T}_\v
+\int\limits_{_{\vec\u\in\Sigma}}
{\colorb{\bs\beta}(\vec\u)}\,{\mathbbm T}_\u
\Bigg]\;
\left.{\colorb\frac{dN_1}{dyd^2\vec\p_\perp}}\right|_{_{\rm LO}}\; .
\end{equation}
In this equation that relates the spectrum at LO and NLO, ${\mathbbm
  T}_\u$ is a functional derivative with respect to the classical field,
and the functions ${\cal G}$ and $\beta$ can in principle be evaluated
analytically ($\Sigma$ is a space-like surface used to set the initial
condition for the classical field). The next step to justify the
universality of the logarithms of energy is to prove that \cite{GelisLV3}
\begin{eqnarray}
    \frac{1}{2}\int\limits_{_{\vec\u,\vec\v\in\Sigma}}
{\colorb{\cal G}(\vec\u,\vec\v)}\,{\mathbbm T}_\u{\mathbbm T}_\v
+\int\limits_{_{\vec\u\in\Sigma}}
{\colorb{\bs\beta}(\vec\u)}\,{\mathbbm T}_\u
= \log\left(\Lambda^+\right)\,{\colorb{\cal H}_1}
    +
    \log\left(\Lambda^-\right)\,{\colorb{\cal H}_2}
    +
    \mbox{terms w/o logs}\; ,
\end{eqnarray}
where ${\cal H}_{1,2}$ are the JIMWLK Hamiltonians of the two
projectiles and $\Lambda^\pm$ the cutoffs that separate the fields
from the fast sources in the CGC effective theory. A striking property
of this formula is that the logarithms of $\Lambda^+$ (cutoff for the
right-moving sources) and of $\Lambda^-$ (cutoff for the left-moving
sources) do not mix: their coefficients are operators that depend only
on $J_1$ or on $J_2$ respectively, but not both. It is this property
that ensures that we can factorize these logarithms into distributions
$W_1[J_1]$ and $W_2[J_2]$ that obey the JIMWLK equation,
\begin{equation}
\Lambda\frac{\partial W[J]}{\partial \Lambda} = {\cal H}\,W[J] \; .
\end{equation}
After averaging over these distributions of sources, the inclusive
gluon spectrum at leading logarithmic accuracy reads
\begin{equation}
\left<\frac{dN_1}{dy d^2\vec\p_\perp}\right>_{_{\rm LLog}}
=
\int 
\big[D{\colora J_{_1}}\,D{\colorb J_{_2}}\big]
\;
{\colora W_1\big[J_{_1}\big]}\;
{\colorb W_2\big[J_{_2}\big]}
\;
\underbrace{\frac{dN_1[J_{1,2}]}{dy d^2\vec\p_\perp}}_{\rm for\ fixed\ J_{1,2}}
\Bigg|_{_{\rm LO}}
\; .
\end{equation}
It is this factorization formula that establishes a link between heavy
ion collisions and DIS on a nucleus, since the same distribution
$W[J]$ appears in both reactions.

The same reasoning can be applied to multi-gluon spectra, for which a
similar factorization holds \cite{GelisLV4},
\begin{eqnarray}
  \left<
    \frac{dN_n}{dy_1d^2\vec\p_{1\perp}\cdots dy_nd^2\vec\p_{n\perp}}
  \right>_{_{\rm LLog}}
  \!\!=  \int 
  \big[D{\colora J_{_1}}\,D{\colorb J_{_2}}\big]
  \;
  {\colora W_1\big[J_{_1}\big]}\;
  {\colorb W_2\big[J_{_2}\big]}
  \;
  \frac{dN_1[J_{1,2}]}{dy_1d^2\vec\p_{1\perp}}\Bigg|_{_{\rm LO}}\!\!\!\!
  \cdots
  \frac{dN_1[J_{1,2}]}{dy_nd^2\vec\p_{n\perp}}\Bigg|_{_{\rm LO}}\! .
\label{eq:corr}
\end{eqnarray}
Since its integrand is fully factorized, this formula tells us that,
at leading logarithmic accuracy, all the correlations between gluons
separated in rapidity come from the evolution of the distributions
$W[J]$ (in other words, these correlations are a property of the
pre-collision initial state). On the basis of this formula, we expect
long range correlations, over separations $\Delta y\sim \alpha_s^{-1}$
(since this is the change in rapidity that one needs in order to have
an appreciable change in the distribution $W[J]$).

\begin{figure}[h]
\begin{center}
\resizebox*{!}{3.7cm}{\includegraphics{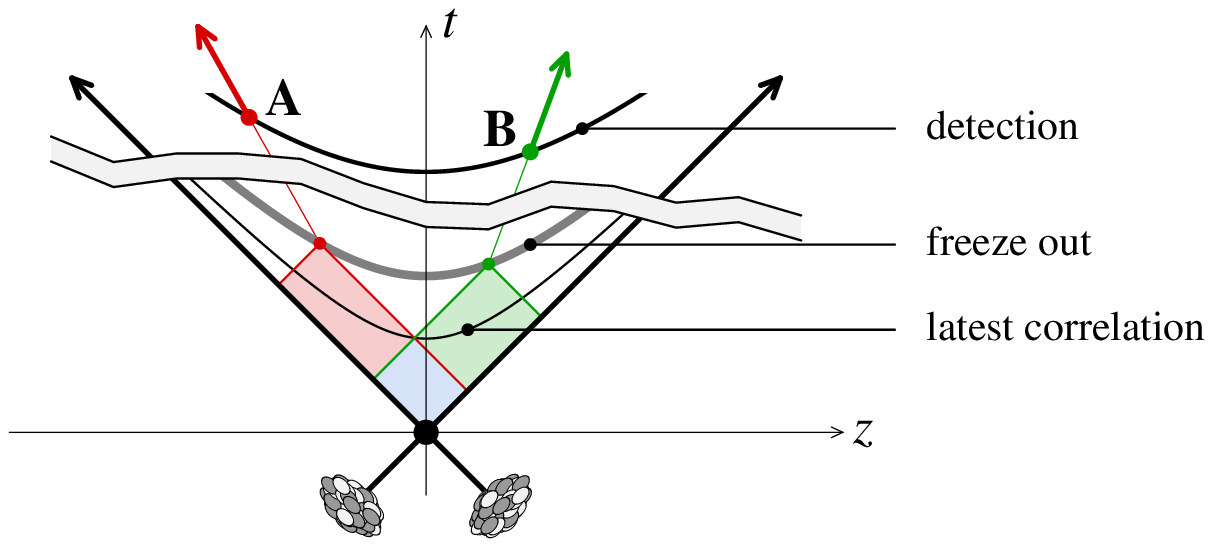}}
\hfil
\resizebox*{!}{4.5cm}{\includegraphics{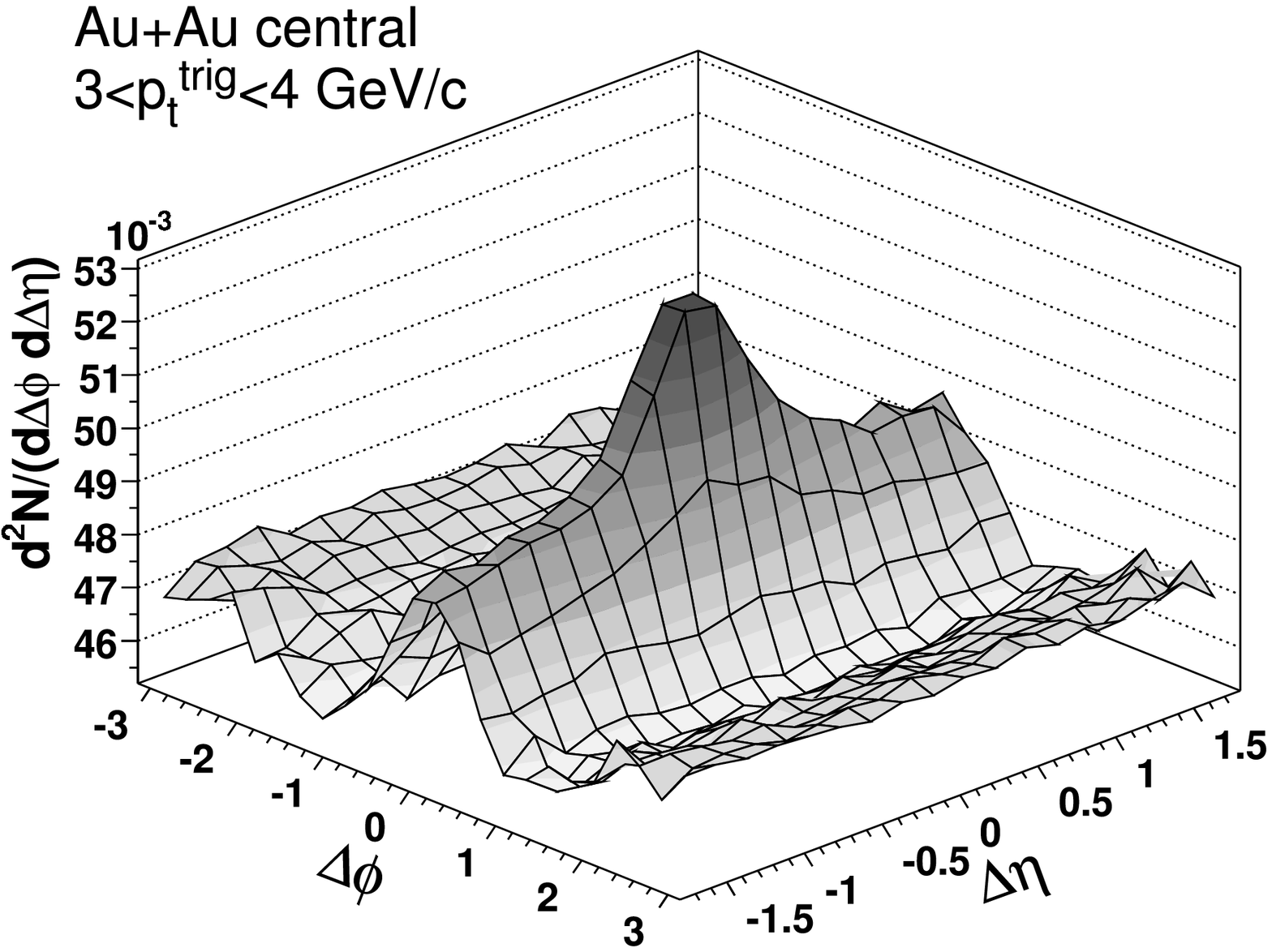}}
\end{center}
\caption{\label{fig:horizon}Left: causal connection between two
  particles at a large rapidity separation. Right: 2-hadron
  correlation in central heavy ion collisions (from \cite{Abelea1}).}
\end{figure}

These rapidity correlations are particularly interesting
experimentally in order to probe the early stages of heavy ion
collisions, because a simple causality argument \cite{DumitDGJL1} (see
the figure \ref{fig:horizon}, left) indicates that long range rapidity
correlations cannot be produced long after the collision.
More precisely, if two particles are detected with a rapidity
separation $\Delta y$, any process inducing a correlation between them
must have taken place at a time $\tau$ smaller than $\tau_{\rm freeze\
  out}\; e^{-|\Delta y| / 2}$, which is a fairly stringent upper limit
due to the exponential factor. As it turns out, such correlations have
been observed in central heavy ion collisions (figure
\ref{fig:horizon}, right). These 2-hadron correlations extend quite
far in rapidity \cite{DumitDGJL1}, but are quite narrow in azimuthal
angle. Note that the formula (\ref{eq:corr}), while it predicts long
range correlations in rapidity, does not lead to any appreciable
collimation of the correlation in azimuthal angle. This is because it
applies to the multi-gluon spectrum only at very short times after the
collision. Subsequently, these gluons will be transported outwards by
the hydrodynamical flow, and it is this radial flow that explains the
azimuthal collimation seen experimentally \cite{Volos1}.

\section{Final state evolution and thermalization}
When discussing the flux tube pattern that emerges at early times in
the classical solution of the Yang-Mills equations, we mentioned in
passing that this implies a negative longitudinal pressure.  To be
more definite, let us compare the energy-momentum tensor obtained at
$\tau=0^+$ in the CGC framework at LO, with the tensor one would have
in ideal hydrodynamics,
\begin{equation}
    {\colord      T^{\mu\nu}_{_{\rm CGC, LO}}(\tau=0^+)
          =
          \begin{pmatrix}
            \epsilon & & & \\
            &\epsilon && \\
            &&\epsilon&\\
            &&&{\colorb-{\epsilon}}\\
          \end{pmatrix}}
\quad,\quad
      { T^{\mu\nu}_{\rm ideal\ hydro}
          =
          \begin{pmatrix}
            \epsilon & & & \\
            &p&& \\
            &&p&\\
            &&&p\\
          \end{pmatrix}}
\; .
\end{equation}
With viscous hydrodynamics, one may cope with a certain degree of
asymmetry between the transverse and longitudinal pressures, but not
to the extent required to accommodate a negative longitudinal pressure.
\begin{figure}[h]
\begin{center}
\resizebox*{!}{4.5cm}{\rotatebox{-90}{\includegraphics{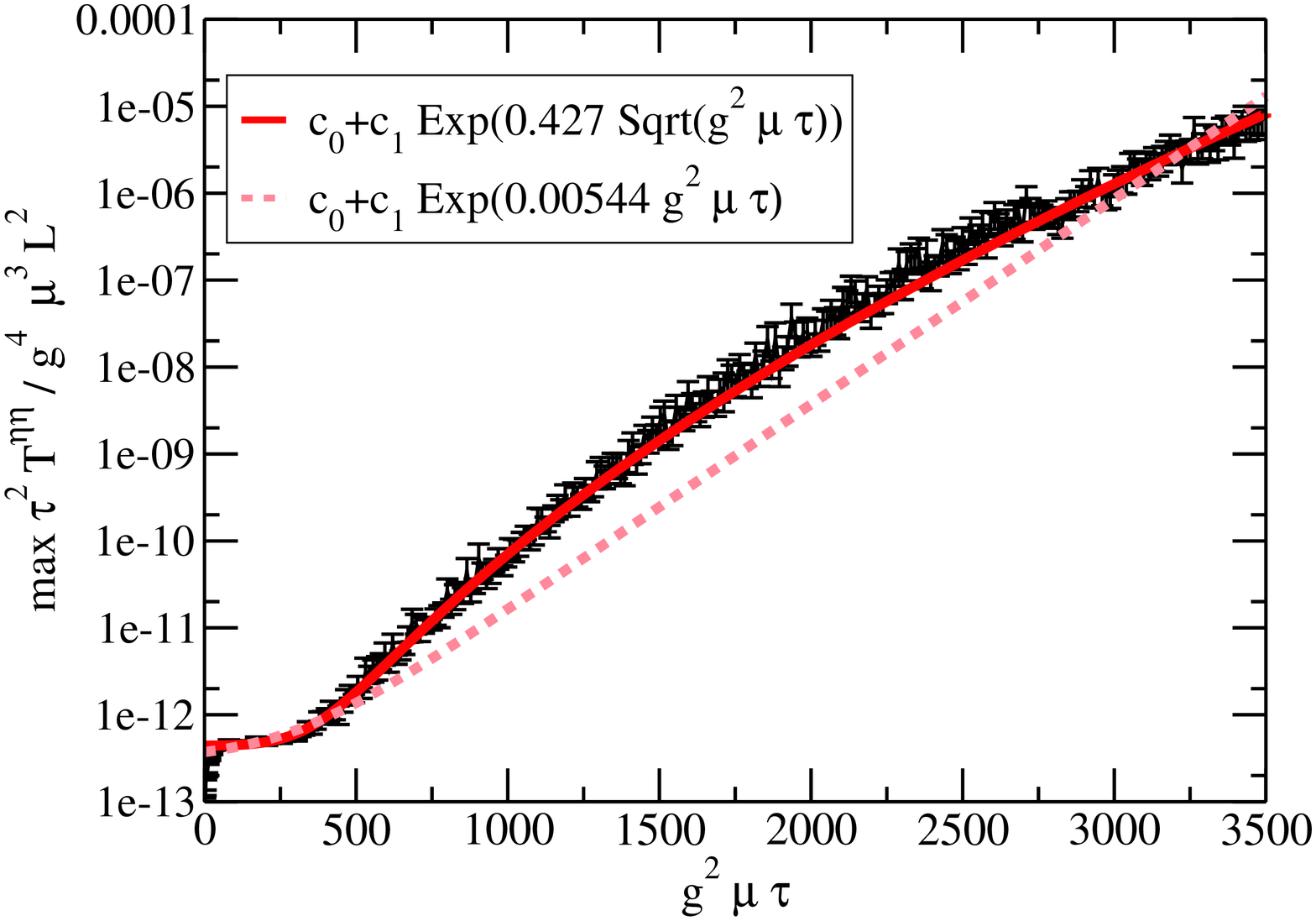}}}
\hfil
\resizebox*{!}{4.5cm}{\rotatebox{-90}{\includegraphics{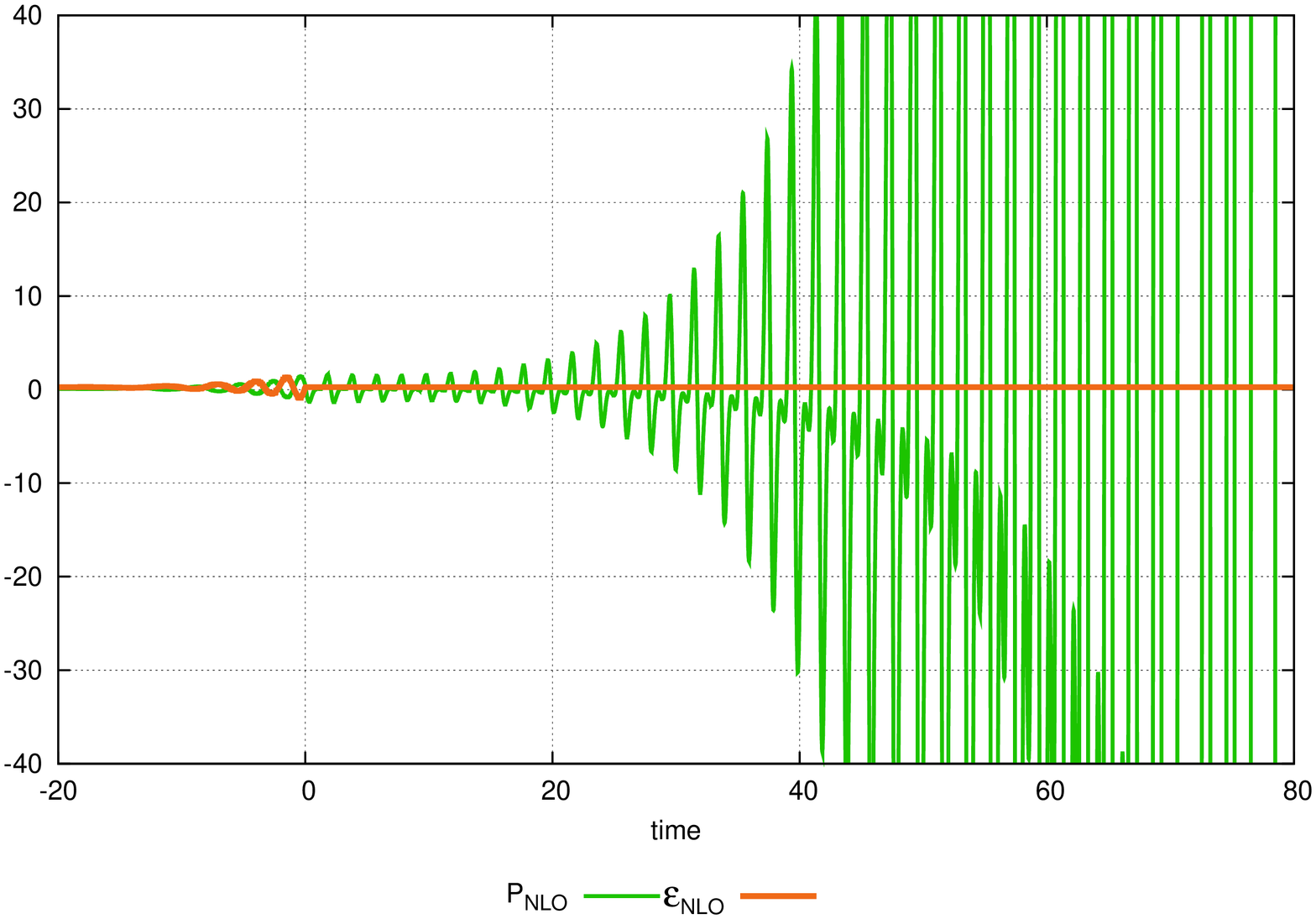}}}
\end{center}
\caption{\label{fig:insta}Left: instabilities in classical Yang-Mills
  equations (from \cite{RomatV1}). Right: energy density and pressure
  at NLO in a toy scalar model.}
\end{figure}
However, it has also been known for a while that the LO result cannot
be the final answer, due to the existence of instabilities in
asymmetric systems of classical color fields
\cite{RomatV1,HeinzHLMM1}.  In the CGC, this issue appears in the form
of unstable modes in the classical Yang-Mills equations: when the
initial condition is perturbed by a rapidity dependent fluctuation,
the solution may diverge exponentially from the unperturbed solution
(see the figure \ref{fig:insta}, left). These unstable modes first
appear in the energy-momentum tensor at NLO, where they lead to
secular divergences in the pressure, and thus to a complete breakdown
of the expansion in powers of $\alpha_s$. It would therefore be highly
desirable to perform a resummation that cures these divergences.

In order to explore these ideas while avoiding most of the technical
complications specific to a gauge theory, we have considered a much
simpler $\phi^4$ scalar toy model that exhibits similar problems. It
Lagrangian is given by
\begin{equation}
{\cal L}=\frac{1}{2}(\partial_\alpha\phi)^2-V(\phi)+J\phi\quad,\qquad
V(\phi)=\frac{g^2}{4!}\phi^4\; ,\quad
  J\propto\theta(-x^0)\; .
\end{equation}
Like a Yang-Mills theory, this model is scale invariant in $3+1$
dimensions (i.e. the only dimensionful scale in our model is brought
by the classical source $J$). In order to mimic a collision, we turn
off the source at $x^0>0$ (thus, the sole purpose of the source is to
initialize the classical fields at $x^0=0$). The reason why this model
is a good playground to study the CGC instabilities is that it also
has unstable modes due to parametric resonance. The dramatic effect
of these instabilities is shown in the right panel of the figure
\ref{fig:insta}, where one sees that the NLO correction to the
pressure explodes exponentially with time.

In order to improve the late time behavior of the pressure, we need a
resummation that restores the non-linearities in the unstable modes
(these modes are linearized when one does an expansion in powers of
$g^2$ at fixed order). A good starting point to find such a
resummation is the formula that relates the LO and NLO:
\begin{equation}
{\colorb T^{\mu\nu}_{_{\rm NLO}}}
=
\Bigg[
\frac{1}{2}\int\limits_{_{\vec\u,\vec\v\in\Sigma}}
{\colorb{\cal G}(\vec\u,\vec\v)}\,{\mathbbm T}_\u{\mathbbm T}_\v
+\int\limits_{_{\vec\u\in\Sigma}}
{\colorb{\bs\beta}(\vec\u)}\,{\mathbbm T}_\u
\Bigg]\;
 {\colord T^{\mu\nu}_{_{\rm LO}}}\; .
\end{equation}
By exponentiating the operator in the square brackets, 
\begin{equation}
{\colorb T^{\mu\nu}_{_{\rm resummed}}}
=
{\colora\exp}\Bigg[
\frac{1}{2}\int\limits_{_{\vec\u,\vec\v\in\Sigma}}
{\colorb{\cal G}(\vec\u,\vec\v)}\,{\mathbbm T}_\u{\mathbbm T}_\v
+\int\limits_{_{\vec\u\in\Sigma}}
{\colorb{\bs\beta}(\vec\u)}\,{\mathbbm T}_\u
\Bigg]\;
 {\colord T^{\mu\nu}_{_{\rm LO}}}\; ,
\end{equation}
it is clear that one obtains in full the LO and NLO contributions,
plus a subset of the higher order terms,
\begin{equation}
T^{\mu\nu}_{_{\rm resummed}}\sim{\colorb\frac{1}{g^2}}\;
\Big[\underbrace{c_0+c_1\,{\colord g^2}}_{\mbox{fully}}+\underbrace{c_2\,{\colord g^4}+\cdots}_{\mbox{partly}}\Big]\; .
\end{equation}
This exponentiation can be justified as the resummation that
picks at each order in $g^2$ the fastest growing unstable terms. Let
us now show that this resummation eliminates the secular divergence
encountered at NLO, by noticing that
\begin{eqnarray}
&&
  {\colora\exp}\Bigg[
\frac{1}{2}\int\limits_{_{\vec\u,\vec\v}}
\!{\colorb{\cal G}(\vec\u,\vec\v)}\,{\mathbbm T}_\u{\mathbbm T}_\v
+\int\limits_{_{\vec\u}}
{\colorb{\bs\beta}(\vec\u)}\,{\mathbbm T}_\u
\Bigg]\,T^{\mu\nu}_{_{\rm LO}}[\varphi_{\rm init}]
\nonumber\\
&&
=
\int[D{\colord\alpha}]\,
\exp\Bigg[-\frac{1}{2}\int\limits_{_{\vec\u,\vec\v}}
\!{\cal G}^{^{-1}}\!(\vec\u,\vec\v)
{\colord\alpha(\vec\u)}{\colord\alpha(\vec\v)}\Bigg]\,
T^{\mu\nu}_{_{\rm LO}}[\varphi_{\rm init}+{\colorb \beta}+{\colord\alpha}]\; .
\end{eqnarray}
In words, this resummation amounts to a Gaussian smearing of the
initial classical field in the LO calculation. Since the LO result is
free of any secular divergence, this resummed result is finite at all
times as well. A similar resummation has in fact been proposed in
other areas before, such as inflationary cosmology \cite{Son1} and
Bose-Einstein condensation \cite{NorriBG1}. This Gaussian ensemble of
initial fields can also be interpreted as a pure quantum mechanical
coherent state, because it has the minimal extension required by the
uncertainty principle, shared evenly by the fields and their conjugate
momenta.

The other virtue of this resummation is that it is fairly easy to
implement numerically, in order to check what effect it has on the
evolution of the energy-momentum tensor.
\begin{figure}[h]
\begin{center}
\resizebox*{!}{4.5cm}{\rotatebox{-90}{\includegraphics{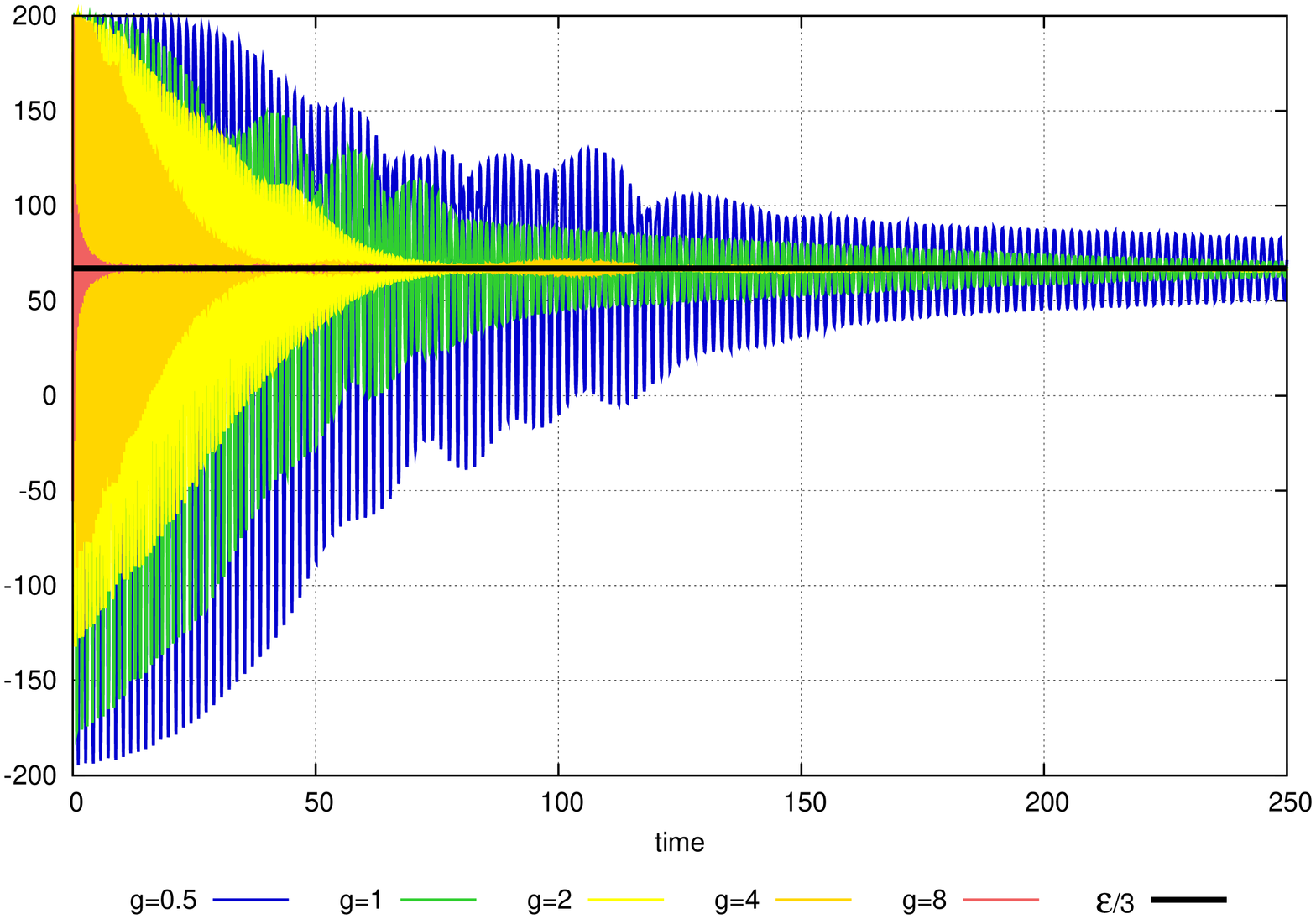}}}
\hfil
\resizebox*{!}{4.5cm}{\rotatebox{-90}{\includegraphics{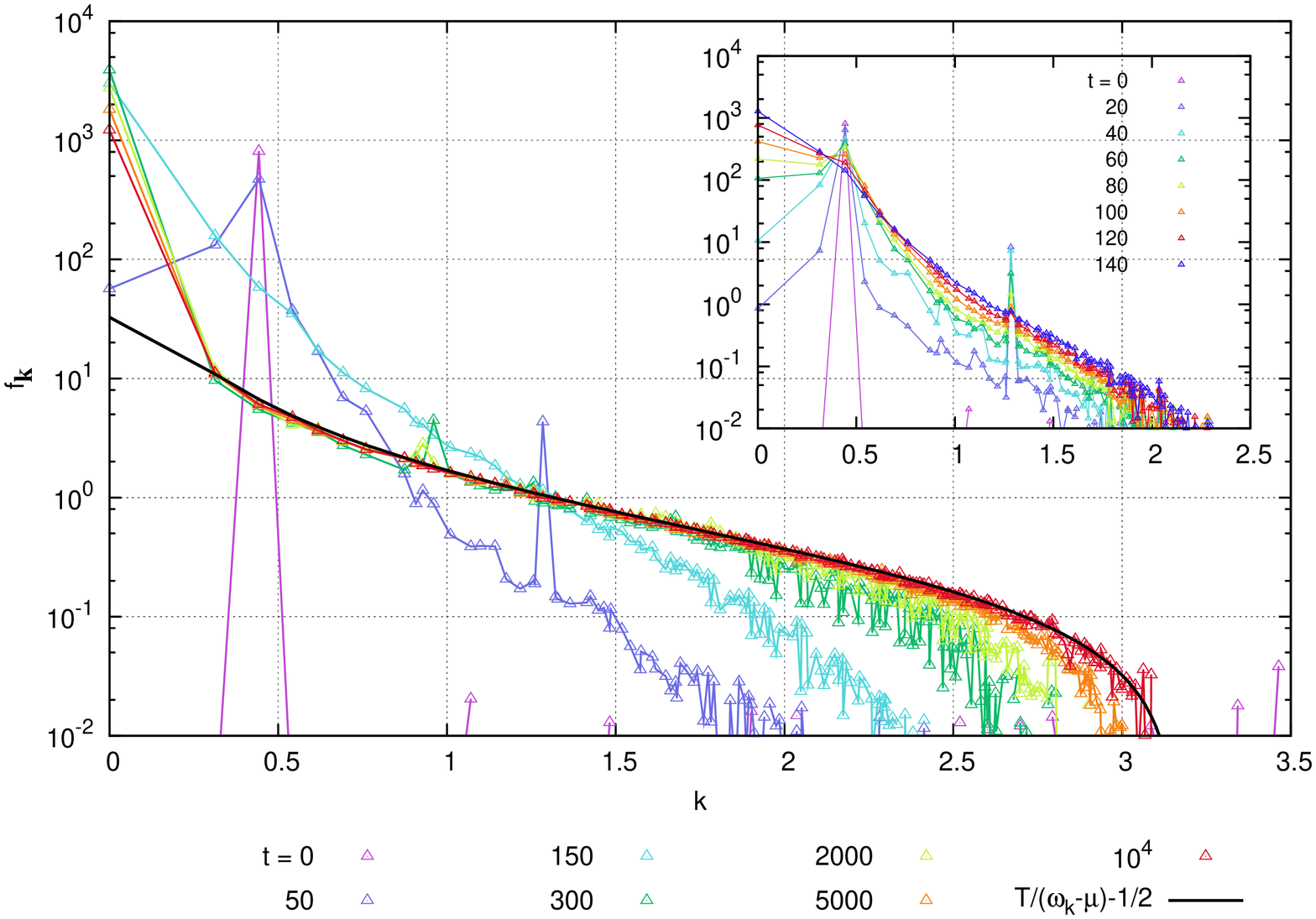}}}
\end{center}
\caption{\label{fig:phi4}Left: energy density and pressure after
  resummation, for various values of the coupling $g$. Right:
  occupation number at various times.}
\end{figure}
This computation merely amounts to solving classical field equations
of motion (the non-linear Klein-Gordon equation in the case of this
scalar toy model) repeatedly with initial conditions drawn randomly
from a Gaussian ensemble~\cite{DusliEGV1}. When this procedure is
applied to the energy-momentum tensor, one finds a result which is
very different from what we got in a strict NLO calculation (see the
figure \ref{fig:phi4}, left). Now, not only the pressure does not
diverge at late times, but its oscillations are damped and eventually
one has $p=\epsilon/3$, which is nothing but the equilibrium equation
of state of a scale invariant system in three spatial dimensions. Note
also that the relaxation time of the pressure decreases with
increasing values of the coupling constant. In a simplified study
where only the zero momentum fluctuations are included, it is possible
to see analytically that the relaxation of the pressure is due to the
loss of coherence of the initial wavefunction under the effect of
nonlinear interactions.  One can also compute the occupation number in
this system and its time evolution (figure \ref{fig:phi4}, right). One
sees that it evolves to an equilibrium distribution, but on
time-scales that are longer than those associated to decoherence.
Interestingly, when the initial condition corresponds to a large
energy density carried by soft modes, a Bose-Einstein condensate forms
in the system, and survives for a very long time if the coupling is
weak. Indeed, at weak coupling the inelastic interactions that could
get rid of the excess of particles in the condensate are quite
suppressed compared to the elastic processes that tend to populate the
condensate. Although this study seems promising, it would be extremely
interesting to see how much of these features survive in a
longitudinally expanding system --as is the case in the geometry of a
collision--, with gauge fields rather than scalar fields
(see \cite{BlaizGLMV1} for recent works).

\ack{I would like to thank the organizers of the Rutherford Centennial
  Conference on Nuclear Physics for their kind invitation, and my
  collaborators J.-P. Blaizot, K. Dusling, T. Epelbaum, K. Fukushima,
  T. Lappi, J. Liao, L. McLerran and R. Venugopalan.}

\section*{References}
\bibliographystyle{unsrt}

\end{document}